\documentclass[%
 aip,
 amsmath,amssymb,
 reprint,%
]{revtex4-1}

\usepackage{graphicx}% Include figure files
\usepackage{dcolumn}% Align table columns on decimal point
\usepackage{bm}% bold math
\usepackage[utf8]{inputenc}
\usepackage[T1]{fontenc}
\usepackage{mathptmx}
\usepackage{etoolbox}

\usepackage[usenames,dvipsnames]{color}
\usepackage{easyReview}
\usepackage{subfigure}
\usepackage{graphics}

\makeatletter
\def\@email#1#2{%
\endgroup
\patchcmd{\titleblock@produce}
{\frontmatter@RRAPformat}
{\frontmatter@RRAPformat{\produce@RRAP{*#1\href{mailto:#2}{#2}}}\frontmatter@RRAPformat}
{}{}
}%
\makeatother

\begin{document}

\preprint{AIP/123-QED}

\title[Sample title]{Metastability in individual magnetic vortices}
% Force line breaks with \\
\author{D. Garc\'{i}a-Pons}
\affiliation{ 
	Instituto de Nanociencia y Materiales de Arag\'{o}n (INMA), CSIC-Universidad de Zaragoza,
	Zaragoza, Spain
}%
\author{J. P\'{e}rez-Bail\'{o}n}
\affiliation{ 
	Instituto de Nanociencia y Materiales de Arag\'{o}n (INMA), CSIC-Universidad de Zaragoza,
	Zaragoza, Spain
}%
\author{A. M\'{e}ndiz}
\affiliation{ 
	Instituto de Nanociencia y Materiales de Arag\'{o}n (INMA), CSIC-Universidad de Zaragoza,
	Zaragoza, Spain
}%

\author{V. J\'{u}lvez}
\affiliation{ 
	Instituto de Nanociencia y Materiales de Arag\'{o}n (INMA), CSIC-Universidad de Zaragoza,
	Zaragoza, Spain
}%

\author{M. Hack}
\affiliation{Physikalisches Institut,  Center for Quantum Science (CQ) and LISA${^+}$,Universit\"at T\"ubingen, Auf der Morgenstelle 14, 72076 T\"ubingen, Germany}

\author{K. Wurster}
\affiliation{Physikalisches Institut,  Center for Quantum Science (CQ) and LISA${^+}$,Universit\"at T\"ubingen, Auf der Morgenstelle 14, 72076 T\"ubingen, Germany}

\author{R. Kleiner}
\affiliation{Physikalisches Institut,  Center for Quantum Science (CQ) and LISA${^+}$,Universit\"at T\"ubingen, Auf der Morgenstelle 14, 72076 T\"ubingen, Germany}

\author{D. Koelle}
\affiliation{Physikalisches Institut,  Center for Quantum Science (CQ) and LISA${^+}$,Universit\"at T\"ubingen, Auf der Morgenstelle 14, 72076 T\"ubingen, Germany}

\author{M. J. Mart\'{i}nez-P\'{e}rez}%
\email{pemar@unizar.es}
\affiliation{ 
	Instituto de Nanociencia y Materiales de Arag\'{o}n (INMA), CSIC-Universidad de Zaragoza,
	Zaragoza, Spain
}%

\date{\today}% It is always \today, today,
%  but any date may be explicitly specified

\begin{abstract}
	Magnetic nanoparticles play a crucial role in different fields such as biomedicine or information and quantum technologies. These applications require nanoparticles with a single, well-defined energy minimum, free of metastable states, and characterized by narrow switching field distributions. Here, we demonstrate that high-transition-temperature nanoSQUIDs can be successfully applied to the characterization of individual nanodiscs hosting magnetic vortices. We present measurements performed under varying temperature and external magnetic field, revealing signatures of ubiquitous, multiple metastable configurations. We also demonstrate that metastability can be reduced by introducing an intended asymmetry. NanoSQUID measurements can be applied to optimize the fabrication of on-demand spin-texture states, such as degenerated vortices or particles with fixed circulation and deterministic and narrow switching probabilities.
\end{abstract}

\maketitle

%%%%%%%%%%%%%%%%
%%%% INICIO %%%%
%%%%%%%%%%%%%%%%

Magnetic vortices \cite{Shinjo2000,Wachowiak2002} arise in confined ferromagnets as a result of the balance between magnetostatic and exchange energies: magnetization curls in plane with defined circulation, forming a vortex core pointing out of plane with up or down polarity. Magnetic vortices have been extensively investigated for applications in spintronics due to their stable non-uniform magnetization (e.g., as logic units\cite{Haldar2016,Jung2012,Pigeau2010,Bowden2010} or to induce superconducting correlations in ferromagnets through the proximity effect\cite{Fermin2023,Lahabi2017}) and due to their rich dynamical behavior in the sub-GHz and GHz range (e.g., as spin-torque nano-oscillators\cite{Shreya2023,Boehnert2023,Pribiag2007} or as spin-wave emitters\cite{Wintz2016}). Despite this amount of interesting physics, quantum properties of magnetic vortices arising at very low temperatures are largely unknown. Looking at quantum technologies, magnetic textures confined in nanoscopic magnets are attracting much attention in the growing field of cavity magnonics\cite{Jiang2023,Rameshti2022,LachanceQuirion2019}. Domain walls, vortices and skyrmions are topological solitons that can be used to encode quantum states \cite{Trif2024,Pan2024,Psaroudaki2023,Khan2021,Liensberger2021}. Magnetic vortices have also been recently proposed as nanoscopic electron paramagnetic resonance sensors capable of reaching single spin sensitivity \cite{GonzalezGutierrez2024}. Additionally, vortices could potentially serve to increase the weak interaction between superconducting microcircuits and spin qubits, even achieving the challenging regime of strong spin-photon coupling \cite{MartinezPerez2018}. For the development of these applications, it is important to obtain nanomagnets with well-defined spin-texture ground states and deterministic and narrow switching distributions, while minimizing metastability as much as possible\cite{Guslienko2024,FernandezPacheco2017}.

In this work, we show the characterization of individual spin textures stabilized in permalloy (Py, Ni$_{80}$Fe$_{20}$) discs. For this, we use nanoscopic Superconducting Quantum Interference Devices\cite{MartinezPerez2017,Granata2016} (nanoSQUIDs) based on the high-transition-temperature and high-critical-field cuprate superconductor YBa$_2$Cu$_3$O$_7$ (YBCO), operative at variable temperature and under sweeping magnetic fields\cite{MartinezPerez2016,Schwarz2015,Schwarz2013}. On the one hand, we identify the nucleation, displacement and annihilation of the vortex core. On the other, we can also distinguish between clockwise (CW) and counterclockwise (CCW) circulation states. Even more important, the occurrence of different metastable states with similar magnetic configuration at certain temperature ranges is experimentally accessible.

NanoSQUID fabrication and operation is described elsewhere\cite{MartinezPerez2016,Schwarz2015,Schwarz2013,Nagel2010}.
In short: a thin film of YBCO is grown epitaxially using pulsed laser deposition on a SrTiO$_{3}$ (STO) [001] bicrystal substrate, with a 24° grain-boundary misorientation. This structural boundary translates into a YBCO grain boundary Josephson junction (GBJ) \cite{Hilgenkamp2002}. After prepatterning by photolithograpy and Ar ion milling, nanoSQUIDs are patterned by Focused Ion Beam (FIB) milling with 30-keV Ga ions, defining a superconducting loop intersected by two GBJs. Additionally, an evaporated Au layer on top of the YBCO film is used as a resistive shunt to ensure nonhysteretic current-voltage characteristics of the GBJs and to minimize Ga implantation into the YBCO film and excessive heating during FIB milling.

%%%%% Fig.1 %%%%%%%%%%%%%%%%%%%%%%%%%%%%%%
\begin{figure}[t]
	\includegraphics[width=0.99\columnwidth]{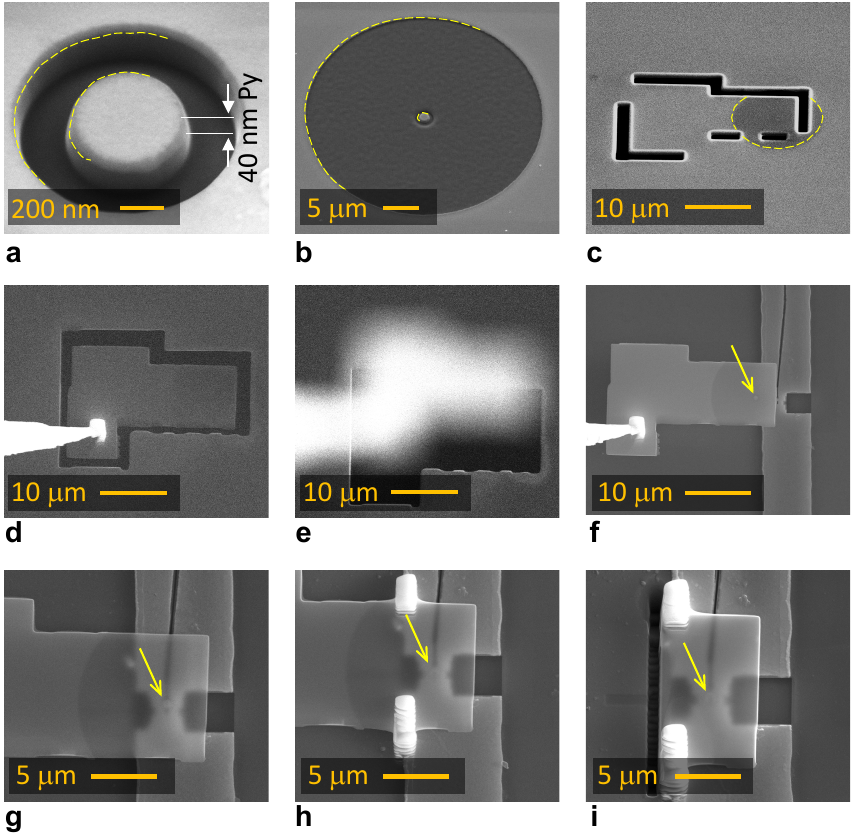}
	\caption{Scanning electron microscopy (SEM) images showing the disc fabrication and transport (see text). The border between the Py film and the Si$_{3}$N$_{4}$ membrane is marked with a dashed yellow line in panels a-c. The Py disc is highlighted by a yellow arrow in panels f-i. }
	\label{fig:fabricacion}
\end{figure}
%%%%% Fig.1 %%%%%%%%%%%%%%%%%%%%%%%%%%%%%%

\begin{figure}[t]
    \includegraphics[width=0.90\columnwidth]{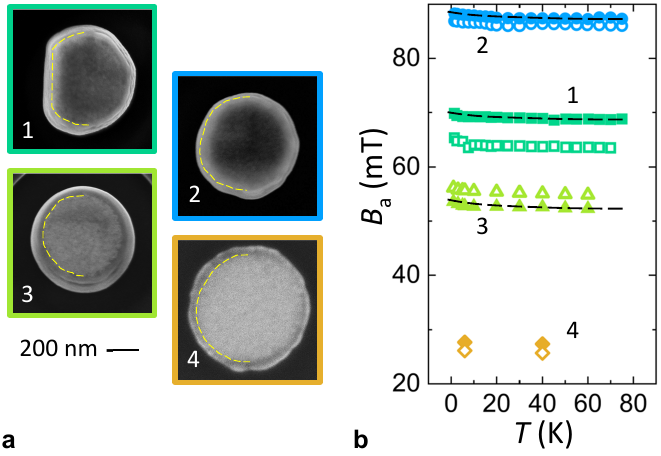}
    \caption{a: SEM images of the samples.  Yellow dashed lines highlight the perimeter of the  Py discs over the Si$_{3}$N$_{4}$ membranes. b: Average positive annihilation fields measured for full loops ($B_{\rm{a}}^{\rm{+full}}$, filled scatter) and semi-loops ($B_{\rm{a}}^{\rm{+}}$, empty scatter). Dashed lines are a guide to the eye to highlight the temperature dependence. }
	\label{fig:discos}
\end{figure}

Fabrication and transport of the Py discs is described in the following. A 40-nm-thick Py film is deposited by e-beam evaporation onto a 200-nm-thick free-standing silicon nitride membrane (PELCO\copyright \; support films for transmission electron microscopy with lateral size of $100$ $\text{\textmu m}$ $\times$ $100$ $\text{\textmu m}$). The disc perimeter is defined by Ga-FIB milling using a low ion current of 10 pA to maximize resolution. Note that Ga ions also partially etch the Si$_{3}$N$_{4}$ layer underneath the Py film (Figure \ref{fig:fabricacion}a). A larger area of the Py film is then removed by FIB-milling with a higher current of 280 pA (panel b).  The entire sample is flipped upside down, and a palette is patterned by Ga-FIB milling (panel c). The Omniprobe needle is attached using a small (focused electron beam induced) deposit of Pt (panel d). The palette is then completely removed and lifted from the remaining silicon nitride membrane (panel e). Subsequently, the palette is approached to the sensor by combining electron and ion beam imaging to precisely control the distance between the palette and the nanoSQUID (panels f and g). The disc position is chosen to optimize the inductive coupling to the nanoSQUID loop\cite{MartinezPerez2018a,MartinezPerez2016}. Finally, two small Pt deposits are used to fix the palette position (panel h), and the remaining part of the palette is removed to eliminate any excess Py film (panel i). Results presented here are obtained with four different Py nanodiscs shown in Figure \ref{fig:discos}a. Disc1 has a maximum diameter of $600$ nm and a flat side that breaks symmetry, whereas disc2 (diameter of $580$ nm), disc3 (diameter of $490$ nm) and disc4 (diameter of $650$ nm) are nominally perfectly circular.

Measurements are performed in a variable temperature inset ($1.4$ K $< T < 80$ K) with a superconducting magnet. The nanoSQUID sensor can be aligned with respect to the external magnetic field $\vec B$ by means of a rotator. In this way, no magnetic flux is coupled to the Josephson junctions, neither to the nanoSQUID (operated in flux-locked loop mode). Hysteresis loops are obtained  by sweeping $B$ while monitoring the SQUID output voltage. The latter is proportional to the total magnetic flux $\Phi$ coupled to the nanoSQUID by the stray magnetic field from the Py disc, i.e., this signal is proportional to the magnetization. 

\begin{figure*}[t]
	\includegraphics[width=0.7\textwidth]{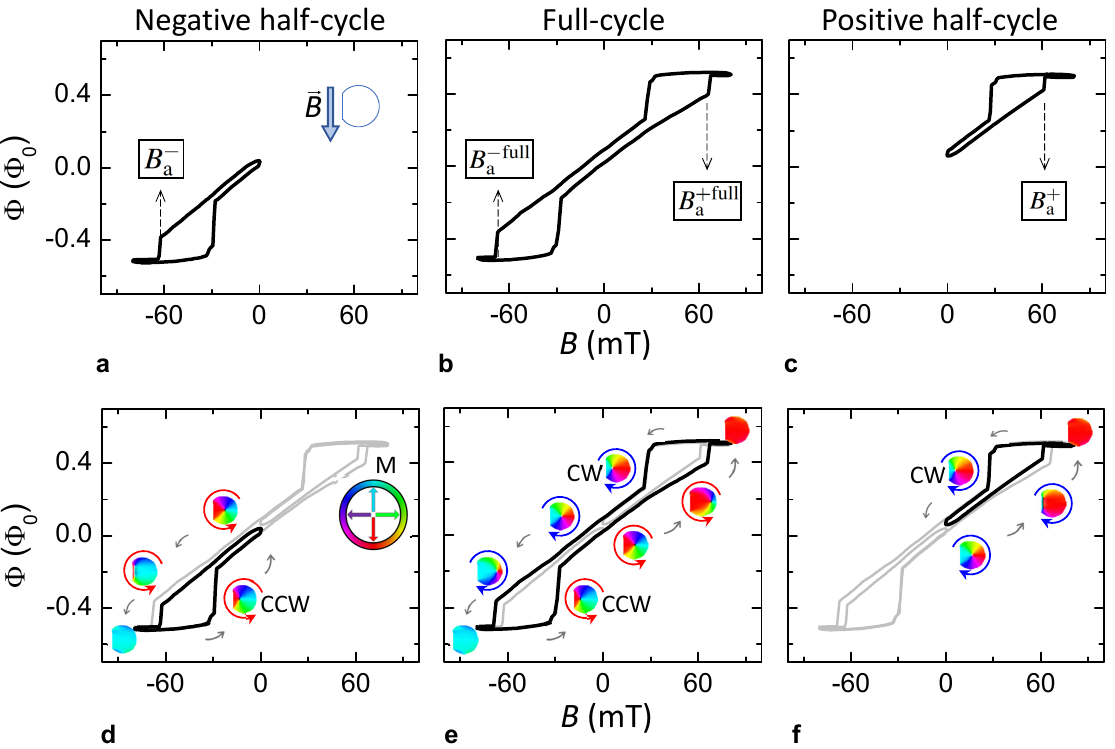}
	\caption{Experimental full and half magnetization cycles $\Phi(B)$ measured for disc1 at $T=70$ K. The field is always applied parallel to the flat side of the particle (see inset in panel a). Vertical axes are given in units of flux coupled to the nanoSQUID loop ($\Phi_0$ is the flux quantum). This is proportional to the nanoSQUID output voltage and, therefore, to the sample magnetization. Annihilation fields $B_{\rm{a}}^{\rm{-}} $, $B_{\rm{a}}^{\rm{-full}} $, $B_{\rm{a}}^{\rm{+full}} $ and $B_{\rm{a}}^{\rm{+}} $ are indicated in panels a-c by vertical dashed arrows. Panels d–f show each hysteresis loop (black curve) alongside the others (grey curves) to highlight the differences between full- and half-cycles. The magnetization configuration is simulated with Mumax and shown at certain field values following the sequence indicated by the small grey arrows (see the magnetization legend in panel d).  Red and blue circular arrows highlight the sense of circulation for CCW and CW directions, respectively. }
	\label{fig:histeresis}
\end{figure*}

Figure \ref{fig:histeresis}b shows a typical hysteresis loop measured at $T~=~70$ K for disc1, with $\vec B$ parallel to the flat side (as shown in the inset of panel a). At $B=0$ the disc is in a vortex state, which produces a very small stray field and, therefore, no net flux coupled to the SQUID. When $B$ is raised, the vortex is displaced perpendicularly to the magnetic field, yielding a linearly increasing flux coupled to the SQUID. At small $B$, the vortex displacement is reversible but, when the field exceeds a certain threshold,  the vortex is expelled out of the disc. This results in the sudden collapse of the magnetization into a saturated state at the annihilation field $B_{\rm a}$. If the magnetic field is reduced from saturation, the vortex state arises again at the nucleation field (typically different from $B_{\rm a}$), resulting in two hysteretic lobes representative for vortex magnetization reversal\cite{Cowburn1999}.

Vortex nucleation is a complex phenomenon, often preceded by the formation of curved magnetization configurations or even two-vortex states\cite{Guslienko2001}. %Additionally, vortex displacement  induced by an applied magnetic field is frequently affected by imperfections or pinning sites,  
This behavior usually leads to the observation of multiple steps in the hysteresis curve, that complicate analysis \cite{MartinezPerez2020}. For this reason, we will focus exclusively on the phenomenon of vortex annihilation and the experimentally observed annihilation fields. We perform three types of measurements, summarized in Figure \ref{fig:histeresis}: (a) half-loops from zero to  negative annihilation field $B_{\rm{a}}^{\rm{-}}$; (b) full-loops showing transitions to both positive and negative saturation states at $B_{\rm{a}}^{\rm{+full}}$ and $B_{\rm{a}}^{\rm{-full}}$, respectively; and (c) half-loops from zero to the positive saturation field, where vortex expulsion occurs at $B_{\rm{a}}^{\rm{+}}$.  %Notice that the vortex core is  annihilated at slightly different values of the applied magnetic field depending on the path followed. 
Measurements are repeated for all discs at varying temperatures. The positive annihilation fields for half- and full-cycles is determined  and shown in Figure \ref{fig:discos}b. As indicated by the dashed lines (used as a guide for the eye), $B_{\rm a}(T)$ follows a logarithmic dependence\cite{Garg1995,Gunther1994,Kurkijaervi1972}. This behavior is well-known and expected from a thermally activated process over an energy barrier, such as vortex annihilation \cite{MartinezPerez2020,Melkov2013,Kakazei2011,Mihajlovic2010,Davis2010,Burgess2010}. What is notable is the fact that  $B_{\rm{a}}^{\rm{+}} \neq B_{\rm{a}}^{\rm{+full}}$ for all discs and across the entire temperature range accessible experimentally. The same behavior is emphasized in panels d-f in Figure \ref{fig:histeresis}.

%%%%% Fig.2 %%%%%%%%%%%%%%%%%%%%%%%%%%%%%%
%%%%% Fig.2 %%%%%%%%%%%%%%%%%%%%%%%%%%%%%%

%%%%% Fig.3 %%%%%%%%%%%%%%%%%%%%%%%%%%%%%%
%%%%% Fig.3 %%%%%%%%%%%%%%%%%%%%%%%%%%%%%%

%%%%% Fig.4 %%%%%%%%%%%%%%%%%%%%%%%%%%%%%%
\begin{figure*}
    \centering
    \includegraphics[width=0.8\textwidth]{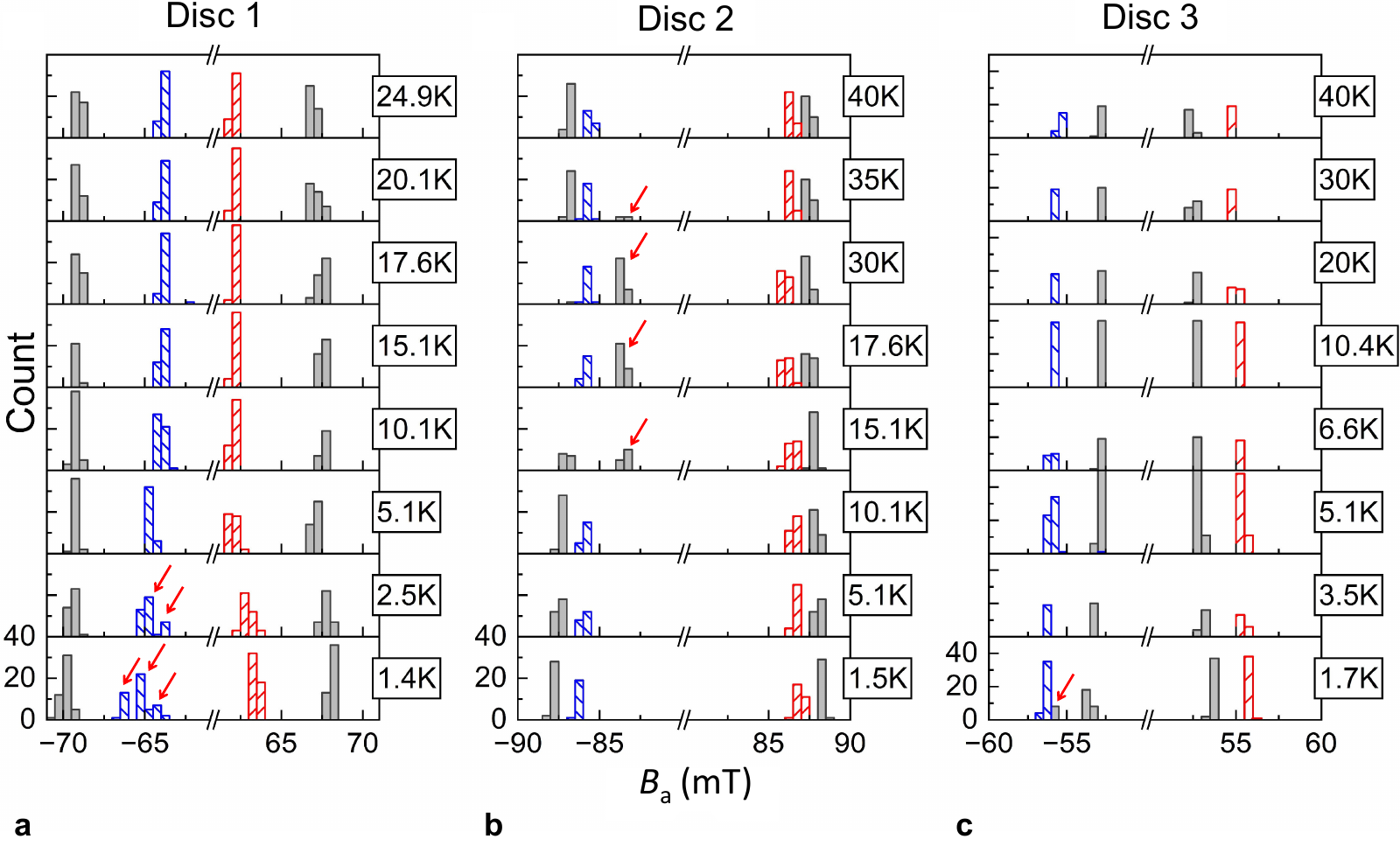}
    \caption{Histograms of annihilation fields  obtained from 40-50 repetitions at different temperatures given in the legend. Filled bars represent data extracted from full magnetization cycles ($B_{\rm{a}}^{\rm{-full}}$ and $B_{\rm{a}}^{\rm{+full}}$); while hatched blue bars ($B_{\rm{a}}^{\rm{-}}$) account for negative semi-loops, and hatched red bars ($B_{\rm{a}}^{\rm{+}}$) correspond to positive semi-loops. Small red arrows highlight signatures of metastability in all discs (see text). %$B_{\rm{a}}^{\rm{-}}$ data corresponding to disc2 at $15.1$ K are absent due to technical problems during measurements.
    }
    \label{fig:multipleStates}
\end{figure*}
%%%%% Fig.4 %%%%%%%%%%%%%%%%%%%%%%%%%%%%%%

To understand this, let us first focus on disc1, with a flat side parallel to the externally  applied magnetic field (see Figure \ref{fig:histeresis}a). In perfectly circular discs, both vortex circulation and polarity states are degenerated. The degeneracy can be easily broken by introducing some asymmetry, such as a flat side\cite{Schneider2001}. Under such circumstances, the probability of the vortex nucleating near the flat side will be higher due to the larger demagnetizing field.  When this happens at positive $B$ (Figure \ref{fig:histeresis}e and f), this process will always yield a CW rotating vortex. In contrast, vortex nucleation at negative fields will yield a CCW configuration (Figure \ref{fig:histeresis}d and r). Interestingly, the circulation will, in turn, fix the direction of the vortex displacement under sweeping $B$, determining the side where vortex expulsion occurs. The annihilation fields will be slightly different depending if expulsion takes place from the flat or the round side of the disc. This is to say, annihilation fields will depend on the circulation, providing an easy method to measure the vortex's direction of rotation.

This effect has been experimentally observed, by measuring half- and full-hysteresis loops in arrays\cite{Dumas2009} and pairs of nanodiscs\cite{Kimura2007}. Full cycles result in vortex nucleation and annihilation at opposite sides of the disc, whereas half cycles cause the vortex to nucleate and annihilate at nearby locations. This is to say, vortex annihilation in the half-cycles will always occur near the flat side, where its demagnetization energy is higher and, therefore, will take place at lower magnetic fields. On the other hand, annihilation in the full-cycles will take place near the curved side, with lower demagnetization energy, and thus will happen at slightly higher fields. In other words, $|B_{\rm{a}}^{\rm{-}}| < |B_{\rm{a}}^{\rm{-full}}|$ and $B_{\rm{a}}^{\rm{+}}<B_{\rm{a}}^{\rm{+full}}$. These phenomena are fully visible in the measurements shown in Figure \ref{fig:histeresis} and summarized in Figure \ref{fig:discos}b for variable temperature. To better illustrate this effect we also perform micromagnetic simulations using MuMax3\cite{Exl2014,Vansteenkiste2014} considering the shape of disc1 as determined from the SEM images and typical material parameters for Py \cite{InfoPy}. Results from the simulations are shown in Figure \ref{fig:histeresis}d-f.

Notably, disc2, disc3 and disc4 also exhibit different annihilation fields for full- and half-cycles (Figure \ref{fig:discos}b). This is probably due to the fact that unintended sample imperfections such as deviations from perfect circular shape, edge roughness or inter-grain boundaries can also break the symmetry\cite{Uhlir2013,Im2012}. This might also determine the direction of circulation and, therefore, the magnetization configuration before vortex annihilation. Unlike the disc with a flat side, in these cases we cannot easily predict the position of vortex nucleation and, therefore, the direction of circulation. Disc3 is particularly intriguing because, contrary to intuition,  $|B_{\rm{a}}^{\rm{-}}| > B_{\rm{a}}^{\rm{-full}}$ and $B_{\rm{a}}^{\rm{+}} > B_{\rm{a}}^{\rm{+full}}$. This is interesting since the vortex is typically nucleated close to the region with maximum energy. During the half-cycles, the vortices nucleate and annihilate near the same location, so we would expect $|B_{\rm{a}}^{\rm{-}}| < B_{\rm{a}}^{\rm{-full}}$ and $B_{\rm{a}}^{\rm{+}} < B_{\rm{a}}^{\rm{+full}}$. However, annihilation does not occur from exactly the same magnetization configuration as nucleation. It is possible for the vortex to become trapped in a lower-energy metastable state. When this happens, vortex expulsion is less favourable, resulting in a higher annihilation field. A similar behavior has been observed in asymmetric discs such as disc1 when the external field is applied at intermediate angles with respect to the flat side\cite{Dumas2009}.

As previously discussed, ideal vortex states are four-fold degenerate in circulation and polarity. Imperfections and asymmetries break this degeneracy but also allow for the stabilization of very similar vortex states, either with nearly identical magnetization configurations or with the vortex pinned at slightly different positions \cite{Im2012,Aliev2011}. These differences result in variations in the corresponding annihilation fields. NanoSQUID measurements on individual discs enable us to distinguish the nucleation of these metastable states and their relative probabilities, which depend not only on the sample quality but also on the temperature. To further explore this behavior we repeat each hysteresis measurement 40-50 times and make histograms of annihilation fields for half- and full-cycles. Results corresponding to disc1, disc2 and disc3 at different temperatures are shown in Figure \ref{fig:multipleStates}. 

Disc1 behaves as expected, with $|B_{\rm{a}}^{\rm{-}}|~<~|B_{\rm{a}}^{\rm{-full}}|$ and $B_{\rm{a}}^{\rm{+}}~<~B_{\rm{a}}^{\rm{+full}}$ at all temperatures. On the other hand, $B_{\rm{a}}^{\rm{-}}$ broadens as temperature decreases, even splitting into two (or more) multiple states at the lowest temperatures below $5$ K (highlighted with arrows). This is a signature of metastability. Different annihilation fields result from slightly different magnetization configurations, each occurring with a certain probability. The height and shape of the annihilation histogram depend on the probability distribution of non-identical metastable configurations. We recall that, ideally, the mechanism for vortex expulsion should be equal for both positive and negative half-cycles. However, the fact that $B_{\rm{a}}^{\rm{+}}$  does not exhibit signatures of metastability suggests that the annihilation mechanism is different for CW and CCW configurations.

Disc2 and disc3 are even more complex. For example, below 15 and above 35\,K,  disc2 is characterized by well defined values of $B_{\rm{a}}^{\rm{+full}}$ and $B_{\rm{a}}^{\rm{-full}}$. On the other hand, at temperatures between 15 to 35\,K, $B_{\rm{a}}^{\rm{-full}}$ splits into two distinct states; at 15.1\,K both states share similar probability. Interestingly, one of these two states (highlighted with an arrow) annihilates at fields smaller that the corresponding $B_{\rm{a}}^{\rm{-}}$ (just as disc3). Disc3 exhibits quite narrow distributions of the annihilation fields, always keeping $|B_{\rm{a}}^{\rm{-}}| > |B_{\rm{a}}^{\rm{-full}}|$ and $B_{\rm{a}}^{\rm{+}} > B_{\rm{a}}^{\rm{+full}}$. Interestingly, at the lowest temperature $T=1.7$ K, 
$B_{\rm{a}}^{\rm{-full}}$ also splits into two distinct states, one being comparable to $B_{\rm{a}}^{\rm{-}}$ (highlighted wiht an arrow). This behavior is compatible with the interpretation of multiple metastable states, having energies higher/lower than the original nucleation configuration and therefore yielding distinct annihilation fields. Just as it happened with disc1, metastable states become apparent only for negative $B$ and, therefore, only for one specific circulation.

To conclude, we have demonstrated ultra-sensitive nanoSQUID measurements on individual Py nanodiscs. By sweeping the external magnetic field, we can distinguish the processes of vortex nucleation, displacement, and annihilation. By performing multiple measurements at varying temperatures, we also obtained histograms of vortex annihilation. These reveal signatures of ubiquitous multiple metastable configurations. Although disc2 and disc3 had a quite regular circular shape, they likely exhibit defects that lead to different vortex pinning centers. The introduction of intentional asymmetries leads to much more reliable behavior, as demonstrated by disc1, although the presence of multiple metastable states at low temperatures is still evident.

NanoSQUID characterization is, therefore, a powerful tool to differentiate between non-ideal particles (with multiple metastable states) and "ideal" particles, characterized by a single, well-defined minimum energy state with optimal behavior and narrow switching distributions. This can be applied for any kind of magnetization configuration, including textures or single domain magnetic states. For example, one could apply this technique to improve the fabrication method of nanodiscs, aiming to obtain either asymmetric particles with well-defined circulation direction or vortex states with degenerate circulation and polarity.

%As seen in the SEM images in Figure \ref{fig:discos}, disc2 has a visibly irregular edge, which could explain this behavior. Discs 2 and 3 both show multiplicity of states below 3\,K for negative half-cycle and full cycle measurements, respectively.

%vortex cores can be pinned by  defects trapping the system in the metastable state. \cite{Aliev2011}

\vspace{5mm}

This work is partly funded and supported by the European Research Council (ERC) under the European Union’s Horizon 2020 research and innovation programme (948986 QFaST), the Spanish MCIN/AEI/10.13039/501100011033 and the European Union FEDER through project PID2022-140923NB-C21, the CSIC program for the Spanish Recovery, Transformation and Resilience Plan funded by the Recovery and Resilience Facility of the European Union, established by the Regulation (EU) 2020/2094, the CSIC Research Platform on Quantum Technologies PTI-001, the Arag\'{o}n Regional Government through project QMAD (E09\_23R) and MCIN with funding from European Union NextGenerationEU (PRTR-C17.I1) promoted by the Government of Arag\'{o}n. We also acknowledge support by the COST actions FIT4NANO
(CA19140) and SUPERQUMAP (CA21144).

\section{AUTHOR DECLARATIONS}
The authors have no conflicts to disclose.

\section{AUTHOR CONTRIBUTIONS}

\textbf{D. García-Pons:} Formal Analysis (equal), Investigation (equal), Visualization (equal), Writing/Original Draft Preparation (lead).
\textbf{J. Pérez-Bailón:} Investigation (equal).
\textbf{A. Méndiz:} Investigation (equal).
\textbf{V. Júlvez:} Investigation (equal).
\textbf{M. Hack:} Resources (equal), Investigation (equal).
\textbf{K. Wurster:} Resources (equal), Investigation (equal).
\textbf{R. Kleiner:} Funding Acquisition (equal), Investigation (equal), Methodology (equal), Resources (equal), Supervision (equal).
\textbf{D Koelle:} Funding Acquisition (equal), Investigation (equal), Methodology (equal), Resources (equal), Supervision (equal), Writing/Review \& Editing (equal).
\textbf{M. J. Martínez-Pérez:} Conceptualization (lead), Formal Analysis (lead), Funding Acquisition (equal), Investigation (lead),
Methodology (lead), Resources (equal), Supervision (lead), Validation (lead), 
Visualization (equal), Writing/Review \& Editing (lead)

\section{DATA AVAILABILITY}

The data that support the findings of this study are available from the corresponding author upon reasonable request.

%\section{AUTHOR DECLARATIONS} \subsection{Conflict of Interest} The authors have no conflicts to disclose. \subsection{Author Contributions} \textbf{David Garc\'ia-Pons}: Formal Analysis (equal); Investigation (supporting); Validation (equal); Visualization (equal); Writing - Review \& Editing (equal). \textbf{Jorge P\'erez-Bail\'on}: Investigation (supporting); \textbf{Arturo M\'endiz}: Formal Analysis (supporting); Investigation (supporting); \textbf{Violeta J\'ulvez}: Formal Analysis (supporting); Investigation (supporting); \textbf{Benedikt M\"uller}: Resources (equal). \textbf{Jianxin Lin}: Resources (equal). \textbf{Martin Hack}: Resources (equal). \textbf{J. Ses\'e}: Formal Analysis (supporting). \textbf{Reinhold Kleiner}: Resources (equal). \textbf{Dieter Koelle}: Resources (equal). \textbf{Mar\'ia Jos\'e Mart\'inez-P\'erez}: Conceptualization (lead); Formal Analysis (equal);Funding Acquisition (lead); Investigation (lead); Methodology (lead); Resources (equal); Software (lead); Supervision (lead); Validation (equal); Visualization (equal); Writing - Review \& Editing (equal). \section{DATA AVAILABILITY} The data that support the findings of this study are available from the corresponding author upon reasonable request.

%\nocite{*}
%\printbibliography % Produces the bibliography via BibTeX.
\bibliography{pydiscs}

\end{document}